\documentclass[
    ,final            
  ]
  {aipproc}

\layoutstyle{8x11double}

\begin{document}

\title{The Night before the LHC\\
{\normalsize ---thoughts about expectations in the early stage 
and beyond---}}


\author{Mihoko M. Nojiri}{
  address={IPNS, KEK, Oho 1-1, Tsukuba, Ibaraki, 305-0801,  Japan \\and  }, 
  altaddress={IPMU, Tokyo University, Kashiwano-ha, Kashiwa, 277-8568, Japan} 
}

\begin{abstract}
I review  
 recent developments on the use of $m_{T2}$ variables for 
 SUSY parameter study, which might be useful for  analyses of the data  
 in the early stage of the LHC experiments.  I also discuss some of 
 recent interesting  studies.  
\end{abstract}

\maketitle


\section{The night before....}
{
'Twas the night before Christmas  \\
\hskip 1cm 
when all through the house \\
Not a creature was stirring, not even a mouse; \\
\\
The stockings were hung by the chimney with care\\
In hopes that St. Nicolas soon would be there' \\\\
\hskip 1cm  {\it --- Clement Clarke Moore  \\
\hskip 1.4cm "The Night before Christmas"} }
\\
\\
Although not in its full scale, the LHC is  starting this year.
All parts of the accerelator and the detectors will  be put together with 
care, in hopes the beams soon would be there---
to reveal the nature of elementary particles 
at  TeV scale. 
Theorists are waiting  something new,  {\it  the gift}  (the item may vary 
from the CMSSM minimum to unparticle), and each of you must have your 
special plans for {\it the night before the LHC}.   Most likely,  the era of 
\lq \lq the freedom of model building" will be over within 
a few years. There will be more data, more 
constraints, more handles. But when and how? 

For the case of supersymmetric models and  its family which 
predict new colored particles decaying into SM particles and 
{\rm a  stable new particle}---a dark matter candidate, 
the discovery channels have been  studied intensively.  The stable particles, 
the lightest supersymmetric particle(LSP) in the minimal supersymmetric 
standard model (MSSM),  give a 
large missing momentum to the events, which 
is an important signature of the SUSY events. 
To control the large QCD, $t\bar{t}$, $W$ and $Z$ backgrounds, 
we require large $E_{\rm Tmiss}$, several high $p_T$ 
jets, large effective mass $m_{\rm  eff}$.
The  ATLAS and CMS studies show that 
 squarks and gluino with the mass below  1.2~TeV would be explored
for   $\int dt {\cal L}=1{\rm fb}^{-1}$ at $\sqrt{s}=14$~TeV. 
Note that the integrated luminosity is required to understand the 
detector and the backgrounds from the data, and the number of SUSY events 
at $1{\rm fb}^{-1}$ itself may be  large enough to allow 
some kind of model parameter studies

Because SUSY at the LHC have been discussed 
in this conference series  for  years and years, 
I concentrate  on  recent developments
on the use of $m_{T2}$ variable in the first part of my talk. It 
might be relevant to the early stage analyses of the LHC and has not been 
systematically studied by the experimental groups yet.  At the 
end of my talk I will also cover other interesting 
developments.

\section{The $m_{T2}$ distribution and sparticle mass determination }
\subsection{The Stransverse mass and the LSP mass determination}

An important development since SUSY '07  is 
a  new understanding on the role of  stransverse mass. 
The stransverse mass, especially,  so called a $m_{T2}$ variable 
has been known  for years\cite{Lester:1999tx, Barr:2003rg}. The  $m_{T2}$  can be 
defined when there  are two visible objects with momenta $p^{i}_{\rm vis}$
 and the missing momentum $p_{\rm miss}$  in an event
as follows, 
\begin{equation}\label{one}
m_{T2}={\rm min}\left[ 
 {\rm max} \left(
m_{T}(p^1_{\rm vis},p^{\rm T}_{\rm 1 },m),
m_{T}(p^2_{\rm vis},p^{\rm T}_{\rm 2 },m),
\right)
\right],
\end{equation}
where the minimum must be  taken for the 
test LSP momenta $p_1$ and $p_2$ which 
satisfy following condition, 
\begin{equation}
{\bf p}^T_{\rm miss}= {\bf p}^{\rm T}_{1} +{\bf p}^{\rm T}_{2},
\label{two}
\end{equation}
and $m$ is a test LSP mass. 
This quantity is an extension of the  transverse mass in the hadron 
collider analysis, aiming  for events with two missing massive particles. 
For  sparticle (co-)production, 
the  true LSP momenta can be a trial LSP momenta of Eq. (\ref{two}), therefore, 
$m_{T2}$ is bounded from above, 
\begin{equation}
m_{T2}< {\rm max} (m_1, m_2)  
\end{equation}
where $m_1$ and $m_2$ are the masses of the primary produced 
SUSY particles.  This is because the $m_{T2}$ is defined as the 
 minimum of the {\it maximum} of the two $m_{T}$,  therefore 
 it effectively  takes the minimum of  the  transverse  mass 
 of  the decay products of the heavier sparticle\cite{Nojiri:2008vq}.  

The  $m_{T2}$ distribution 
can be  used  for the determination 
the right handed squark mass ($m_R$)  in 
$jj$+$E_{Tmiss}$ channel\cite{Weiglein:2004hn}. When $m_{R}< m_{\tilde{g}}$, 
$\tilde{q}_R$ dominantly decays into the  lightest neutralino $\tilde{\chi}^0_1$ 
 and a jet, 
therefore $\tilde{q}_R\tilde{q}_R$ production can be tagged by requiring 
two very high $p_T$ jets in the final state.  When $m=m_{LSP}$, the 
endpoint of $m_{T2}$ is equal to $m_{R}$. 
The selected events are populated near the $m_{T2}$ endpoint, therefore 
the SM background is negligible.  The right-handed squark mass 
 $m_R$ is determined with the error of  3\% at SPS1a\cite{Weiglein:2004hn}.

Recently, Cho et al\cite{Cho:2007qv, Cho:2007dh} investigated the $m_{T2}$ variable as a function of 
a  test LSP mass for gluino pair production and the decay $pp\rightarrow \tilde{g}
 \tilde{g}\rightarrow 
jjjj\tilde{\chi}^0_1\tilde{\chi}^0_1$. The 
$m_{T2}$ variable is constructed from two jet pairs
so that $p^{1(2)}_{vis}$ is a momentum of one of the jet pairs
arising from a gluino decay. 
The endpoint of the $m_{T2}$ distribution as a
function of a  test LPS mass  has  a 
kink exactly at $m=m_{LSP}$.  This is  because 
two different sets  of the SUSY events 
 contribute to the endpoint of $m_{T2}$ variable. 
 For $m<m_{LSP}$, the events with  $m_{jj}\sim m^{\rm min}_{jj}$ 
give $m^{\rm max}_{T2}$, while  for $m>m_{LSP}$, 
 the events with  $m_{jj}\sim m^{\rm max}_{ll}$ give 
it, where $m^{\rm min(max)}_{jj}$ is 
the minimum (maximum) jet pair invariant mass 
arising from $\tilde{g}$ decay. 
At $m=m_{LSP}$, the endpoints of both of the events should be 
at $m_{\tilde{g}}$ becuase the true LSP momenta
can be the test momenta that satisfy  Eq.(2).  Thus,  the endpoint becomes a
  function which has a kink at $m=m_{LSP}$.  This means that one can determine 
 the LSP mass from the kink position experimentally. 
 
In \cite{Cho:2007qv,Cho:2007dh},
 it is shown that  the $m_{T2}$  endpoint is sensitive to the both
 LSP  and gluino masses for several model points by explicit MC simulations. 
 They select the 
four highest $p_T$ jets of the events, and divide  them into 
two jet pairs using some distance measure, and regarded 
the jet pair momentum as two visible momenta $p^{i}_{\rm vis}$ 
in Eq. (\ref{one}). 

In the previous analyses,  the endpoint method has been used to   determine the  sparticle  masses. For example, in  the SUSY cascade decay 
$\tilde{q}\rightarrow \chi^0_2\rightarrow \tilde{l}\rightarrow \chi^0_1$,
the sparticle masses are solved analytically from  the endpoints 
of $m_{ll}$, $m_{jl}$ and $m_{jll}$ distributions. 
The lepton channel  is very clean  but 
the branching ratio  is rather small  in  wide region of the parameter 
space. This was a problem in the SUSY parameter determination 
at the LHC---we did not know how to fix the LSP mass 
kinematically when $\tilde{\chi}^0_2\rightarrow \tilde{l}l $ is closed.   
It should be noted that  the  decay  $\tilde{g}\rightarrow jj\chi^0_1$ 
for  the $m_{T2}$ study  
involves 
only jets, and  it has significant branching ratio. The LSP 
mass determination using the $m_{T2}$ kink method 
could be applied in wider parameter region.

\subsection{The inclusive $m_{T2}$}
The proposal in \cite{Cho:2007qv, Cho:2007dh} is still limited to the case that only a single channel 
(either gluino-gluino or squark-squark) contributes to the four jet + missing 
$E_T$ channel, and 
a gluino decays dominantly into $jj\tilde{\chi}^0_1$.  
It is not satisfactory because gluino 
and squark can decay into the channel involving 
multiple jets. The number of 
high $p_T$ jets are  larger than four, and there are no good 
reason to select the first four jets to study $m_{T2}$ distributions
for general MSSM points.  They are also  generally co-produced. 
When $m_{\tilde{g}}\sim m_{\tilde{q}}$, the squark-gluino co-production 
is the dominant part of the colored SUSY particle productions. 

\begin{figure}
  \includegraphics[width=5.4cm]{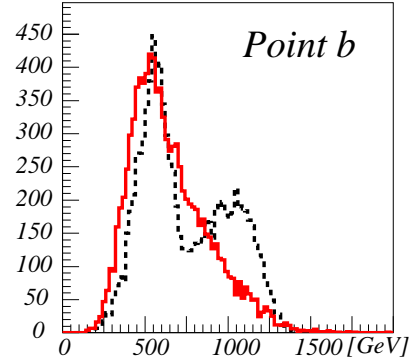}
 \caption{A parton level inclusive  $m_{T2}$ distribution at a model point 
 with $m_{\tilde{q}}=1342$~GeV and $m_{\tilde{g}}=785$~GeV.  
 The solid line correspond to the distribution using the hemisphere algorithm, while the dotted line correspond to those with correct parton assignments.}\label{parton}
 \end{figure}
 \begin{figure}
 \includegraphics[width=6.5cm]{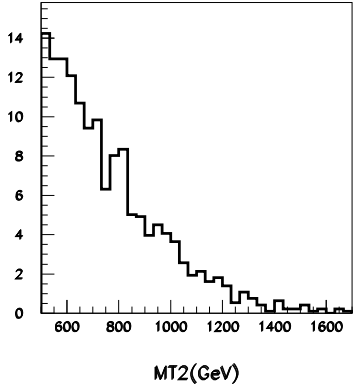}\hskip -1cm
\caption{The inclusive $m_{T2}$ distribution at the model point 
of  Fig. 1 for $\int dt {\cal L}=1$fb$^{-1}$.  
The distribution is after the standard SUSY cuts. 
See \cite{ Nojiri:2008vq}  for details. }\label{jet}
\end{figure}

In \cite{Nojiri:2008hy, Nojiri:2008vq},  we proposed an inclusive definition of 
$m_{T2}$,  which can be calculated 
for  events with any number of jets in the final state.  The quantity is defined 
first by dividing  the jets  into two hemisphere which 
satisfy following conditions, 
\begin{eqnarray}
p^{(i)}_{\rm hemi}&=&\sum_{k\in H_i} p_k
\cr
d(p_k, p^{(i)}_{\rm hemi} )& < &d(p_k, p^{(j)}) 
\ {\rm for}\  k\in H_i\  {\rm  and} \  k\notin H_j
\cr
{\rm  where,}&& 
\cr
d(p_k,p_j)&\equiv& (E_k-\vert p_k\vert \cos \theta_{jk} )
\frac{E_{k}}{(E_j+ E_k)^2}
\end{eqnarray}
The hemisphere axes (momenta)  may be found
 by taking the highest $p_T$ jet $i$ and 
the jet $j$ with ${\rm max} (\Delta R_{ij}  p_{Tj})$ as initial axes.  
Jets are first associated with one of  the initial axes with smaller $d$.  
The new hemisphere momenta is calculated under the assignments, and 
the procedure is iterated until the assignment converges. 
Only the jets with $p_T > 50$~GeV and $\vert \eta\vert <3$ GeV 
are involved in the hemisphere analysis. 

The inclusive  $m_{T2}$ is defined as $m_{T2}$  for 
 $p^{i}_{\rm visi}=p^{(i)}_{\rm hemi}$ in Eq.(\ref{one}).  We study the distribution 
 at the model points with 
 $m_{\tilde{g}}\sim750$~GeV  and $m_{\tilde{q}}$  from $880$~GeV to 1516~GeV
using HERWIG for event generations. 
 These points are within the reach of ATLAS and CMS 
at $\int dt {\cal L}=1$fb$^{-1}$. Some of the mass parameters of the 
model points  in \cite{Nojiri:2008vq} is listed in Table 1.

\begin{table}
\begin{tabular}{|c||c|c|c|c|c|c|}
\hline
& $m_0$& $A_0$ & $m_{\tilde{q}}$& $m_{\tilde{g}}$&  $m_{\rm LSP}$  & $\mu$ \\
\hline
a & 1400& $-1400$&   1516 & 795.7 & 107.9  &  180\\
b & 1200 & $-1200$ &  1342 & 785.0 & 107.4 &  180\\
c & 1100 & $-1100$ & 1257 & 779.5 & 107.1  &  180\\
d & 1000 & $-1000$ &  1175 & 773.2 & 106.8 &  180\\
e &  820 &$-750$ &    1035 & 761.7 & 106.1 &  180\\
f &  600 &$-650$ &  881.0 & 745.4 & 107.8  &  190\\
\hline
\end{tabular}
\caption{Some of the mass parameters of  our model points. 
We take the scalar masses of sfermions and gaugino masses to be universal. 
We tune the higgsino mass parameter $\mu$ by allowing 
non-universal 
GUT scale Higgs masses parameters 
so that  $\Omega h^2 \sim 0.1$.  All mass parameters are given in GeV.  
}\label{model}
 \end{table}
 
 In Fig.~\ref{parton}  we show a  parton level inclusive $ m_{T2}$ distributions for  
a model point with $m_{\tilde{q}}=1342$~GeV and $m_{\tilde{g}}=785$~GeV.  
By using the generator  information, we find that  our hemisphere algorithm 
reconstruct only  $1/4$ of the events without any mis-assignment. 
However, the mis-reconstructed events tend to have 
 smaller $m_{T2}$ value so that the endpoint of 
the  $m_{T2}$  distribution coincides  with
 that of  the correct hemisphere assignment 
 shown in the  dotted line.   We also find the endpoint of  $m^{\rm end }_{T2}$ 
 agree with $m_{\tilde{q}}$ 
 for $m_{\rm test}=m_{LSP}$ for the model points 
 we have studied.  We also show the signal 
$m_{T2}$ distribution using HERWIG for event generation 
and parton shower, with a toy detector simulator using AcerDET
in Fig. \ref{jet}. The obtained distribution  is consistent with 
the parton level distribution in Fig.\ref{parton}.

The merits of the inclusive approach are 1) one can use the 
all events available at the early stage of the experiment 2) 
The end point is least biased.  However there are 
some demerits co-exist as well.  For example,  
the kink of the inclusive $m_{T2}$ distribution is studied in \cite{Nojiri:2008hy}, and  
the result is mixed.  The kink method is most 
effective when the particle contributes to the endpoint 
follow the three body decay, when the difference between 
$m^{\rm min}_{\rm vis}$ and $m^{\rm max}_{\rm vis}$ 
is large.  It is not always guaranteed for the inclusive 
approach. In addition, the distributions show  
tails arising from the contamination of jets 
coming from the radiation from the initial state 
quarks and gluon. 
When enough luminosity is avaiable, 
 selecting clean decay chain would give a better results.

Both the effective mass $m_{\rm eff}\equiv
\sum p^{jet}_T+ E_{\rm Tmiss}$
 and the inclusive $m_{T2}$ involve the missing transverse momentum 
in their definitions. It has been  known that a peak of 
 $m_{\rm eff}$ distribution of SUSY events 
has strong correlation with the parent sparticle masses.  This is because 
the SUSY production occurs dominantly at its threshold,  therefore the sum 
of the $p_T$ of the jets and leptons  reflects the sum of the 
parent sparticle masses.  Obviously, this is a phenomenological relation.  Moreover,  
there are   SM backgrounds  which is not negligible at the peak position. 
Therefore, some  systematics is expected in the  extraction of the peak positions 
of the signal. 

\begin{figure}
\includegraphics[width=6.5cm]{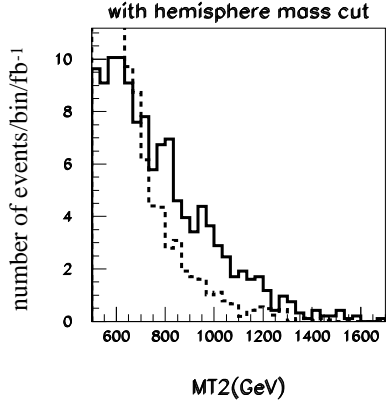}
\caption{The signal(solid) and background(dashed) distributions
at the same model point. Here we apply a  cut 
$m^{1(2)}_{hemi}>200$~GeV.  We produced 50,000 SUSY events 
for this study, and the  distribution is normalized to 1fb$^{-1}$ }\label{sgbg}
\end{figure}

On the other hand,  the inclusive $m_{T2}$ distribution 
has a clear kinematical 
interpretation; the endpoint should {\it  exactly} coincide with the parent 
sparticle mass when $m_{\rm test}=m_{\rm LSP}$.  
Furthermore,  the SM background tends to be suppressed near
 the $m_{T2}$ endpoint.  The background $m_{T2}$  distribution 
are shown with the signal distribution in Fig.~\ref{sgbg}.  The
 $S/N$ ratio near the $m_{T2}$ endpoint is large near the endpoint. 
Here the background distribution contain contributions from 
$t\bar{t} +n$ jets($n\le 2$), $Z^0+n$ jets($n\le 5$), and $W^{\pm}+ n$ jets 
($n\le 4$)  generated and matched using ALPGEN.  

\begin{figure}
\includegraphics[width=6.5cm]{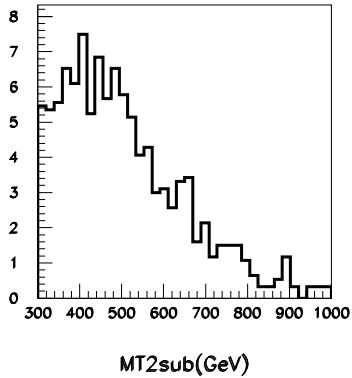}
\caption{The $m^{\rm sub}_{T2}$ 
distributions at the same model point. The endpoint is consistent 
with the input gluino mass $m_{\tilde{g}}=785$~GeV.  
The distribution is normalized to 
1fb$^{-1}$.  }\label{sub}
\end{figure}

\begin{figure}
 \includegraphics[width=5.5cm, angle=90]{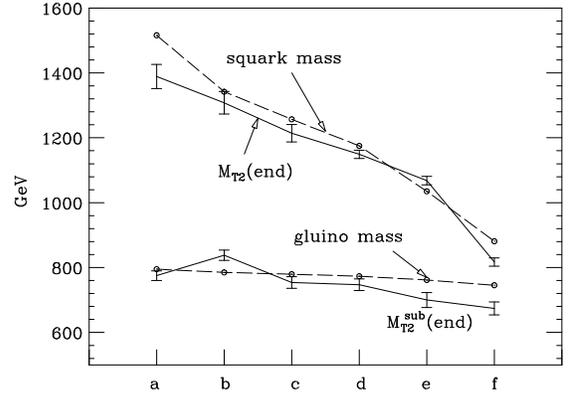}\hskip -1cm
\caption{The fitted $m_{T2}$  and $m^{\rm sub}_{T2}$ 
endpoints(solid lines)  and $m_{\tilde{q}}$ and $m_{\tilde{g}}$ (dashed lines)
at each model point. The bars show statistical errors for 50,000
events. 
}\label{compare}
\end{figure}

The  $m_{T2}$ of  jet subsystem is also useful.  When squarks 
 heavier than a gluino, a squark decay into  
  a high $p_T$ quark and  gluino/neutralinos. We constructed hemispheres 
for the jet systerm without the highest $p_T$ jet and 
calculate the subsystem $m_{T2}$ ($m^{\rm sub }_{T2}$) 
distributions\cite{Nojiri:2008vq}.  
The $m^{\rm sub}_{T2}$ distribution 
has an endpoint consistent with gluino mass when $m_{\tilde{g}}<m_{\tilde{q}}$. 
(See Fig.~\ref{sub}) .  
 This shows that one can extract both squark and gluino masses 
 in the event by looking jet distribution in terms of $m_{T2}$ 
 and $m_{T2}^{sub}$ .  The endpoint values  
of the $m_{T2}$ and $m^{\rm sub}_{T2}$ distributions are obtained by 
a linear fit, and they  are compared with input squark and gluino masses
 in Fig.~\ref{compare}.

 \subsection{Exact relations}
 In the endpoint method,  endpoints of the several 
 distributions are solved to obtain the sparticle masses
 involved in the events. 
Each endpoint gives one constraint among the sparticle masses. 
It is known that 
a decay cascade involving at least  three SUSY cascade decaies 
are  required 
 to determine the all sparticle masses from the endpoints.  
The results in \cite{Cho:2007qv, Cho:2007dh} shows that 
$E_{\rm Tmiss}$ 
constrain the  LSP 
momenta,  and  the spartice 
masses are solved even though there are only two sparticles 
involved in the decays. 
In general,  $p_{\rm Tmiss}$ provides an independent constraint 
to the sparticle mass determination when both of the  
pair produced sparicle decays are  identified.  Several groups studied 
the case involving $\tilde{q}
\rightarrow\tilde{\chi}^0_2\rightarrow\tilde{l}\rightarrow\tilde{\chi}^0_1$. 
By using the exact relation among the visible and missing 
momenta event by event, 
 the errors on the LSP mass is improved by factor of 30\% 
 in \cite{Nojiri:2007pq}

\section{More favorite toys} 
If SUSY particles are found in the early phase of the LHC 
experiments, we would be able to access various decay 
chains of SUSY particles  in the later stage of the experiment. 

The channel involving $\chi^0_i \rightarrow \tilde{l} l$ is experimentally 
clean,  therefore 
search of lepton flavor violation (LFV) in  neutralino cascade decays are
 experimentally promising.  The LFV in SUSY processes might come from 
right-handed neutrino Yukawa couplings. 
Gauge mediation models  may also lead experimentally 
acceptable LFV if the  planck scale  soft masses violate lepton 
flavor\cite{Feng:2007ke}.
 By measuring difference of the two lepton endpoints  $\Delta m_{ll}=
m_{ee}-m_{\mu\mu}$ arising from $\tilde{\chi}^0_2$
$\rightarrow \tilde{l}l $$\rightarrow ll\tilde{\chi}^0_1$, 
one can detect/set the lower limit to  
the slepton mass difference $\Delta m_{l}=m_{\tilde{e}}-m_{\tilde{\mu}}$. 
The $\Delta m_{l}$ is expected to be non-zero in  models with 
non-zero LFV.   The sensitivity to the 
mass difference is recently discussed in \cite{Allanach:2008ib}. 

The masses of the third generation squarks and sleptons are also 
important in  distinguishing SUSY models. They are expected to be lighter 
than the first and second generation squarks if the SUSY mediation 
scale is near the planck scale. In addition,  stop mass 
and its mixing are  important parameters for Higgs mass 
radiative corrections.  

 The sbottom and 
stop masses  may be obtained by studying the 
gluino decays $\tilde{g}\rightarrow \tilde{t}t, \tilde{b}b$. 
The $\tilde{g} \rightarrow bb \tilde{\chi}^0_2$ channel
has been studied and it has been shown that 
the lighter sbottom mass can be obtained from 
$bbll$ channel. 
a recent ATLAS full simulation study confirms the earlier fast simulation 
study\cite{Hisano:2003qu}  on the 
$\tilde{g}\rightarrow \tilde{t}t \rightarrow t b \tilde{\chi}^{\pm}$ 
reconstruction in a model point. The hadronic top decay is reconstructed 
correctly and the $m_{tb}$ endpoint is seen\cite{giacomo}.

Finally, there appeared  a few interesting attempts to improve 
 reconstruction of the boosted 
$W$ and $H$ which decays hadronicaly\cite{Butterworth:2007ke, Butterworth:2008iy}. 
The sparticle decays sometimes produce $W$  or $H$ bosons, and it 
decays dominantly into jets. 
One may look for the jet pairs with $m_{\rm pair}$ consistent with 
them, however  there are other jet pairs  or fat jets  within the 
same mass range. 
It is pointed out that these background can be reduced  by looking 
 into the jet substructure. The proposed procedure is 1) 
reconstruct jets  with  
the  $k_T$ algorithm with somewhat large $R(\sim 1)$. 
2) Take a hard jet $i$, break it into two sub-jets $j$ and $k$
by undoing its last stage of clustering. 
3) If there is significant mass drop for the subjet  and 
 $y= \min(p^2_{Tj},p^2_{Tk}) \Delta R^2_{jk}/m^2_{i}> y_{cut}$, then 
 treat it as heavy particle neighborhood. 
 The initial studies show the cut on  $y$ is useful 
 in reducing the background to the heavy particles. 
 Jet substructure or
  jet-jet correlation in SUSY processes and its SM background  
  would be important issues that has not been  fully investigated yet. 

\section{consclusion}
It is likely that  particle physics  will undergo big changes in 
next few years.   LHC will turn on, and we 
are finally able to work toward the theory of 
elementary particles --not just one of the possible beyond the standard 
models. 

 Although the hadron collider is not a perfect 
place to do precise new physics measurements compared with 
high energy $e^+e^-$ colliders,   we now think 
it is possible to do some kind of measurements at the  LHC. 
This {\it  "common sense"} has been  built through  continuous efforts  
to find clear-cut procedures to study the new physics 
in the LHC environment.  The $m_{T2}$ study discussed 
in this review is an example that  finding 
a new analysis method greatly improve our 
understanding  on  
new physics processes. 
It is worth  pushing  this efforts further,  
rather than  judging/estimating  the LHC 
ability  based on 
 our current knowledge, or just worrying about "empty stockings". 
 
At this moment, it is probably wise to discuss 
issues on  the discovery of new physics and the initial data analysis. 
However once some signature of new physics is discovered, we can 
focus our attention on  specific channels which are
sensitive to the nature of new physics sectors. 
In particular, I hope flavor structures 
of  the new physics sector  emerge from the data at the later stage of LHC.  
Let's hope that there will be more experimental plenary talks next year, 
starting to prove the physics beyond the standard model.

\begin{theacknowledgments}
I would like to thank  my collaborators Y. Shimizu, M. Takeuchi 
and K. Sakurai. The work of M.M.N is supported in part by
the Grant-in-Aid for Science Research 
and World Premier International Research Center(WPI) Initiative, 
MEXT, Japan.
\end{theacknowledgments}



\bibliographystyle{aipproc}   


\end{document}